\documentclass{article}
\usepackage{spconf,amsmath,graphicx,bbm,mathtools,multirow,hhline}

\DeclareMathOperator*{\argmax}{arg\,max}

\newlength{\bibitemsep}
\newlength{\bibparskip}
\let\oldthebibliography\thebibliography
\renewcommand\thebibliography[1]{%
  \oldthebibliography{#1}%
  \setlength{\parskip}{\bibitemsep}%
  \setlength{\itemsep}{\bibparskip}%
}

\title{Internal Language Model Training for Domain-Adaptive End-to-End Speech Recognition}
\name{\begin{tabular}{c}Zhong Meng, Naoyuki Kanda, Yashesh Gaur, Sarangarajan Parthasarathy, Eric Sun, \\ Liang Lu, Xie Chen, Jinyu Li, Yifan Gong\end{tabular}}
\address{Microsoft Corporation, Redmond, WA, USA}
\begin{document}
\ninept
\maketitle
\begin{abstract}
The efficacy of external language model (LM) integration with existing end-to-end (E2E) automatic speech recognition (ASR) systems can be improved significantly using the internal language model estimation (ILME) method \cite{meng2021ilme}. In this method, the internal LM score is subtracted from the score obtained by interpolating the E2E score with the external LM score, during inference. 
To improve the ILME-based inference, we propose an internal LM training (ILMT) method to minimize an additional internal LM loss by updating only the E2E model components that affect the internal LM estimation.
ILMT encourages the E2E model to form a standalone LM inside its existing components, without sacrificing ASR accuracy.
After ILMT, the more modular E2E model with matched training and inference criteria enables a more thorough elimination of the source-domain internal LM, and therefore leads to a more effective integration of the target-domain external LM.
Experimented with 30K-hour trained recurrent neural network transducer and attention-based encoder-decoder models, ILMT with ILME-based inference achieves up to 31.5\% and 11.4\% relative word error rate reductions from standard E2E training with Shallow Fusion on out-of-domain LibriSpeech and in-domain Microsoft production test sets, respectively.




\end{abstract}
\section{Introduction}
\label{sec:intro}
End-to-end (E2E) automatic speech recognition (ASR) has achieved state-of-the-art performance by directly mapping speech to word sequences through a single neural network. The most popular E2E models include connectionist temporal classification \cite{graves2006connectionist, hannun2014deep, soltau2016neural, Li18CTCnoOOV}, recurrent neural network transducer (RNN-T) \cite{graves2012sequence, jain2019rnn, sainath2020streaming,  li2020developing}, and attention-based encoder-decoder (AED) models \cite{chorowski2015attention, chan2016listen, chiu2018state, karita2019comparative, Li2020Comparison}.
However, E2E models tend to overfit the audio-transcript pairs in source-domain training data, and suffer from performance degradation when evaluated in a mismatched target domain.
Numerous ideas have been explored to adapt ASR models, such as regularization method \cite{kld_yu, meng2019asa,l2_liao, meng2020lvector, lhuc}, teacher-student learning \cite{li2014learning, meng2018adversarial, manohar2018teacher, meng2019conditional}, transformation method \cite{lhn, tan2015cluster, sc_abdel}, and adversarial learning \cite{grl_shinohara, meng2018speaker, grl_serdyuk, dsn_meng}. Nevertheless, all these methods require audio data for adaptation when applied to E2E models \cite{ochiai2018speaker, meng2019speaker, meng2019domain}.
One promising solution without using audio is to train a language model (LM) with a large amount of text readily available in the target domain and fuse it with the E2E model during inference.
However, an E2E model does not have a modular LM component as in a traditional hybrid system \cite{DNN4ASR-hinton2012}, making external LM integration a challenging task.

Among many approaches proposed for LM integration, Shallow Fusion \cite{hannun2014deep, gulcehre2015on, chorowski2016towards, kim2020improved} is a simple yet effective method where the log probabilities of the E2E model and the LM are linearly interpolated during the inference. Towards a better integration, the Density Ratio method \cite{mcdermott2019density, kanda2016maximum} subtracts the source-domain LM score from the interpolated score of Shallow Fusion, and shows improved performance. Further, as a new type of E2E model, the hybrid autoregressive transducer (HAT) was proposed in \cite{variani2020hybrid} to preserve the modularity of a traditional hybrid system. HAT allows us to estimate the internal LM scores and subtract them from the Shallow Fusion scores for external LM integration.

More recently, we proposed an internal LM estimation (ILME) method in \cite{meng2021ilme} to facilitate the integration of an external LM for any pre-existing E2E models, including RNN-T and AED models, without any additional training.  With ILME-based inference, the internal LM score of an E2E model is estimated by eliminating the contribution of the acoustic encoder, and is then subtracted from the log-linear interpolation between E2E and external LM scores.
However, ILME-based inference in \cite{meng2021ilme} is performed with an ASR model trained to optimize a standard E2E loss by updating all model parameters. The accuracy of the internal LM estimation is not guaranteed when the E2E model is not structured in a way that strictly satisfies the conditions of Proposition 1 in \cite[Appendix A]{variani2020hybrid}. 

To compensate for the mismatch between the E2E training and the ILME-based inference, we propose an \emph{internal LM training (ILMT)} of the E2E model to minimize an additional internal LM loss by updating only the model components engaged in the prediction of internal LM scores during inference. ILMT facilitates the E2E model to form a standalone LM inside its existing components while maintaining ASR accuracy. 
ILMT improves the effectiveness of the ILME-based LM integration with a more modular E2E model and a well-aligned training and inference criterion.
Evaluated with 30 thousand (K)-hour trained RNN-T and AED models, ILMT with ILME-based inference achieves up to 31.5\% and 11.4\% relative word error rate (WER) reductions from Shallow Fusion on cross- and intra-domain evaluations, respectively, far outperforming the reductions with standard E2E training. 

\vspace{-0pt}
\section{Related E2E Methods}
\label{sec:related}
An E2E model predicts the conditional distribution $P(\mathbf{Y} |
\mathbf{X};\theta_\text{E2E})$ of token sequences $\mathbf{Y}=\{y_1, \ldots,
y_U\}$ given a speech-feature sequence $\mathbf{X}=\{\mathbf{x}_1,
\ldots, \mathbf{x}_T\}$ as the input, where $y_u \in \mathcal{V}$ and $\mathbf{x}_t$ is a feature vector at time $t$. $\mathcal{V}$ is the set of all possible output tokens, i.e., word pieces. We insert a start-of-sentence token $y_0=\texttt{<sos>}$ at the beginning of $\mathbf{Y}$.

\subsection{Recurrent Neural Network Transducer (RNN-T)}
The RNN-T model \cite{graves2012sequence} comprises an encoder, a prediction network and a joint network. The encoder maps the input speech features $\mathbf{X}$ to a sequence of hidden states $\mathbf{H}^{\text{enc}} = \{\mathbf{h}^\text{enc}_1, \ldots, \mathbf{h}^\text{enc}_T\}$.
The prediction network is an RNN that takes the embedding vector $\mathbf{e}_{u-1}$ of the previous \emph{non-blank} token $y_{u-1}$ and generates the hidden state $\mathbf{h}^\text{pred}_u$, i.e., $\mathbf{h}^\text{pred}_u = \text{PredictionRNN}(\mathbf{h}^\text{pred}_{u - 1}, \mathbf{e}_{u - 1})$.

The joint network is a feed-forward network that combines the outputs of the encoder and prediction network
to predict the conditional distribution over the next possible token $\tilde{y}_i\in \mathcal{V} \cup \texttt{<b>}$, i.e., 
\begin{align}
    \mathbf{z}_{t_i, u_i} &= W_j \phi(W_e\mathbf{h}^{\text{enc}}_{t_i} + W_p\mathbf{h}^\text{pred}_{u_i} + \mathbf{b}_e + \mathbf{b}_p) + \mathbf{b}_j, \label{eqn:rnnt_logit} \\
    \hspace{-4pt} 
    \left[P(\tilde{y}_i\right.&\left.=v|\mathbf{X}_{1:t_i}, \mathbf{Y}_{0:u_{i - 1}}; \theta_\text{RNNT})\right]_{v\in \mathcal{V} \cup \texttt{<b>}} \hspace{-7pt}= \text{softmax}(\mathbf{z}_{t_i, u_i}), \label{eqn:rnnt_softmax}
\end{align}
where $\texttt{<b>}$ denotes a blank symbol, $\phi$ is a non-linear function, e.g., tanh or ReLU. $W_j$, $W_e$, $W_p$ are weight matrices, and $\mathbf{b}_e$, $\mathbf{b}_p$, $\mathbf{b}_j$ are biases. $\mathbf{z}_{t_i, u_i}$ is a $|\mathcal{V}|+1$ dimensional logit vector. $\tilde{y}_i$ forms a blank-augmented token sequences $\mathbf{\tilde{Y}} = \{\tilde{y}_1, \ldots, \tilde{y}_{T+U}\}$ aligned with the token and feature sequences $\mathbf{Y}$ and $\mathbf{X}$ as $\left(\tilde{y}_i, y_{u_i}, x_{t_i}\right)^{U + T}_{i=1}$, i.e., the index $i$ in $\mathbf{\tilde{Y}}$ is mapped to the index $u_i$ in $\mathbf{Y}$, and the index $t_i$ in $\mathbf{X}$.

The RNN-T loss is computed by marginalizing over all possible blank-augmented token sequences aligned with each reference $\mathbf{Y}$, i.e., $\mathcal{A}(\mathbf{X}, \mathbf{Y})$, on the training corpus.
\begin{align}
    & \mathcal{L}_{\text{RNN-T}}(\theta_\text{RNN-T}) \nonumber \\
    & \hspace{-5pt} = - \hspace{-3pt}\sum_{(\mathbf{X}, \mathbf{Y}) \in \mathcal{D}} \hspace{-4pt} \log \hspace{-9pt} \sum_{\mathbf{\tilde{Y}} \in \mathcal{A}(\mathbf{X}, \mathbf{Y})} \hspace{-2pt} \prod^{T+U}_{i=1} P(\tilde{y}_i|\mathbf{X}_{1:t_i}, \mathbf{Y}_{0:u_{i - 1}}; \theta_\text{RNN-T}). \label{eqn:rnnt_loss}
\end{align}

\subsection{Attention-Based Encoder-Decoder (AED)}
The AED model \cite{chorowski2015attention} consists of an encoder, a decoder and an
attention network. 
The encoder maps a sequence of input speech frames $\mathbf{X}$
into a sequence of hidden states $\mathbf{H}^{\text{enc}}$.
The attention network generates an attention weight for $\mathbf{h}^\text{enc}_t$ at each decoder step $u$,
determining which encoder states should be attended to predict
the output label $y_u$, i.e., $\mathbf{a}_u = \text{AttentionNet}(\mathbf{a}_{u-1}, \mathbf{h}^\text{enc}_t, \mathbf{h}^\text{dec}_u)$,
where $\mathbf{a}_u$ is a vector of attention weights of dimension $T$, and $\mathbf{h}^\text{dec}_u$ is the decoder hidden state.
The context vector $\mathbf{c}_u$ is computed as a linear combination of $\mathbf{H}^{\text{enc}}$ weighted by the attention, i.e., $\mathbf{c}_{u} = \sum_{t = 1}^{T} a_{u,t}\mathbf{h}^\text{enc}_{t}$.

At each step $u$, the decoder RNN takes the sum of the previous token embedding 
$\mathbf{e}_{u-1}$ and the context vector $\mathbf{c}_{u-1}$ as the
input to predict the conditional distribution over $\mathcal{V} \cup \texttt{<eos>}$, i.e., 
\begin{align}
       \mathbf{h}^\text{dec}_u & = \text{DecoderRNN}(\mathbf{h}^\text{dec}_{u-1}, \mathbf{e}_{u-1} + \mathbf{c}_{u-1}), \label{eqn:aed_decoder} \\ 
       \mathbf{z}_u & = W_{d}\mathbf{h}^\text{dec}_u +
       \mathbf{b}_d, \label{eqn:aed_logit} \\
       \left[P(y_u \right. &= \left. v| \mathbf{X}, \mathbf{Y}_{0:u-1};\theta_\text{AED})\right]_{v \in
        \mathcal{V} \cup \texttt{<eos>}} \hspace{-3pt} = \text{softmax}(\mathbf{z}_u), \label{eqn:aed_softmax}
\end{align}
where $\texttt{<eos>}$ is the end-to-sentence token, 
$W_d$ and $\mathbf{b}_d$ are weight matrix and bias, respectively.

The AED loss is obtained as a summation of token sequence posteriors over the training corpus $\mathcal{D}$ as follows
\begin{align}
    \mathcal{L}_{\text{AED}}(\theta_\text{AED}) 
	= \sum_{(\mathbf{X}, \mathbf{Y}) \in \mathcal{D}} \sum_{u = 1}^{U + 1} \log P(y_u |
		\mathbf{X},\mathbf{Y}_{0:u-1};\theta_\text{AED}). \hspace{-2pt}
       \label{eqn:aed_loss}
\end{align}

\section{Internal LM Estimation (ILME)}
\label{sec:ilme}

From audio-transcript training pairs, an E2E model implicitly learns an internal language model (LM) that characterizes the distribution of source-domain training text. The exact computation of internal LM is intractable, but it can be approximated by Proposition 1 in \cite[Appendix A]{variani2020hybrid} 
which suggests that the E2E internal LM $P(y_u|\mathbf{Y}_{0:u-1};\theta^\text{S}_\text{E2E})$ approximately equals to E2E model output $\text{softmax}[J(\mathbf{g}_u)]$ after zeroing out the acoustic embedding $\mathbf{f}_t$ if $P(y_u| \mathbf{X}, \mathbf{Y}_{0:u-1};\theta^\text{S}_\text{E2E})) = \text{softmax}[J(\mathbf{f}_t + \mathbf{g}_u)]$ and $J(\mathbf{f}_t + \mathbf{g}_u) \approx J(\mathbf{f}_t) + J(\mathbf{g}_u)$ are satisfied, where $\mathbf{g}_u$ is a language embedding.


As shown in \cite{meng2021ilme}, the conditional probability of RNN-T internal LM, $P(y_u|\mathbf{Y}_{0:u-1}; \theta_\text{RNNT})$, is estimated as a softmax normalization of the non-blank token logits when the hidden states of the encoder are eliminated from the input of the joint network. 
\begin{align}
    & \mathbf{z}^\text{ILM}_u = W_j\phi(W_p\mathbf{h}^\text{pred}_u + \mathbf{b}_p) + \mathbf{b}_j \label{eqn:logit_ilm}, \\
    & P(y_u|\mathbf{Y}_{0:u-1}; \theta_\text{RNNT}) = \text{softmax}(\mathbf{z}^\text{ILM, NB}_{u}), \label{eqn:rnnt_cond_ilm}
\end{align}
where $\mathbf{z}^\text{ILM}_u$ is a $(|\mathcal{V}| + 1)$-dimensional vector with a designated logit for the blank token $\texttt{<b>}$, $\mathbf{z}^\text{ILM, NB}_u$ is a logit vector of dimension $|\mathcal{V}|$ created by taking out the blank logit from $\mathbf{z}^\text{ILM}_u$. 
Without the encoder input, the RNN-T is completely driven by the prediction and joint networks with the token sequence $\mathbf{Y}$ as the only input. 




Similarly, \cite{meng2021ilme} has also shown that the conditional probability of the AED internal LM is estimated by the decoder output after zeroing out the context vector, i.e,
\begin{align}
    & P(y_u|\mathbf{Y}_{0:u-1}; \theta_\text{AED}) \nonumber \\
    & = \text{softmax}\left[W_d \cdot \text{DecoderRNN}(\mathbf{h}^\text{dec}_{u-1}, \mathbf{e}_{u-1}) + \mathbf{b}_d \right]. \label{eqn:aed_cond_ilm}
\end{align}
Without the context vector, AED is entirely driven by the decoder with the token sequence $\mathbf{Y}$ as the only input, acting exactly the same as an RNN-LM. 

During ILME-based inference \cite{meng2021ilme}, we subtract the log of the internal LM probability $P(\mathbf{Y}; \theta_\text{E2E})$ from the log-linear combination between the conditional probability of E2E model and the external LM probability $P(\mathbf{Y}; \theta_\text{LM})$, and search for the optimal token sequence $\hat{\mathbf{Y}}$ as follows via a left-to-right beam search.
\begin{align}
    \hat{\mathbf{Y}} = \argmax_{\mathbf{Y}} & \left[\log P(\mathbf{Y}|\mathbf{X}; \theta_\text{E2E}) + \lambda_E \log P(\mathbf{Y}; \theta_\text{LM}) \right. \nonumber \\
                     & \left. \quad - \lambda_I \log P(\mathbf{Y}; \theta_\text{E2E}) \right],
\end{align}
where $\lambda_E$ and $\lambda_I$ are the weights for external and internal LMs, respectively.

\section{Internal LM Training of E2E Models}
In standard E2E training, an internal LM is implicitly learned to minimize the E2E loss by updating all parameters of the E2E model. However, during ILME-based inference, only a part of the E2E model contributes to the prediction of the internal LM scores. The estimation of the internal LM scores is not accurate when the conditions of Proposition 1 in \cite[Appendix A]{variani2020hybrid} is not strictly satisfied by the E2E model. 

In this work, we propose an \emph{internal LM training} of the E2E model to mitigate the mismatch between the E2E training and the ILME-based inference. Through the standard E2E training, the decoder of an AED or the prediction and joint networks of an RNN-T acts as an acoustically-conditioned LM that takes both the token and acoustic embeddings as the input to predict the conditional probability of the next token. From Eqs. \eqref{eqn:rnnt_cond_ilm} and \eqref{eqn:aed_cond_ilm}, the internal LM scores are estimated entirely by the acoustically-conditioned LM of an E2E model during ILME-based inference. Therefore, the goal of ILMT is to encourage the acoustically-conditioned LM of an E2E model to also behave like a standalone internal LM, without sacrificing ASR accuracy.
To achieve that, we jointly minimize an internal LM loss together with the standard E2E loss during ILMT.

The internal LM loss of an RNN-T model is obtained by summing up the negative log probabilities of the internal LM over the training corpus $\mathcal{D}$ as follows
\begin{align}
    \hspace{-2pt}\mathcal{L}_{\text{ILM}}(\theta_\text{pred}, \theta_\text{joint}) 
    = -\sum_{\mathbf{Y} \in \mathcal{D}} \sum^{U}_{u=1}\log P(y_u|\mathbf{Y}_{0:u-1};\theta_\text{pred}, \theta_\text{joint}). \label{eqn:rnnt_ilm_loss}
\end{align}
Note that, from Eqs. \eqref{eqn:logit_ilm} and \eqref{eqn:rnnt_cond_ilm}, the RNN-T internal LM loss is conditioned only on the parameters of the prediction and joint networks, $\theta_\text{pred}$ and $\theta_\text{joint}$. 
For RNN-T, the ILMT loss is constructed as a weighted sum of the RNN-T loss in Eq. \eqref{eqn:rnnt_loss} and the ILM loss below 
\begin{align}
    \mathcal{L}_{\text{ILMT}}(\theta_\text{RNN-T}) = \mathcal{L}_{\text{RNN-T}}(\theta_\text{RNN-T})
    + \alpha \mathcal{L}_{\text{ILM}}(\theta_\text{pred}, \theta_\text{joint}), \label{eqn:ilmt_rnnt}
\end{align}
where $\alpha$ is the weight of the internal LM loss. By minimizing the RNN-T ILMT loss, we maximize the internal LM probability of the E2E training transcripts by updating only the prediction and joint networks while maximizing the conditional probability of the training transcripts given input speech by updating the entire RNN-T. 

The internal LM loss of AED is formulated as a summation of negative log probabilities of the internal LM over training corpus $\mathcal{D}$
\begin{align}
    & \mathcal{L}_{\text{ILM}}(\theta_\text{dec}) 
    = -\sum_{\mathbf{Y} \in \mathcal{D}} \sum^{U + 1}_{u=1}\log P(y_u|\mathbf{Y}_{0:u-1};\theta_\text{dec}). \label{eqn:aed_ilm_loss}
\end{align}
Note that, from Eq. \eqref{eqn:aed_cond_ilm}, the AED internal LM loss is conditioned only on the parameters of the decoder $\theta_\text{dec}$. 
For AED, the ILMT loss is computed as a weighted sum of the AED loss and the ILM loss below 
\begin{align}
    \hspace{-4pt}\mathcal{L}_{\text{ILMT}}(\theta_\text{AED}) \hspace{-1pt} = \hspace{-1pt} \mathcal{L}_{\text{AED}}(\theta_\text{AED}) \hspace{-1pt} + \hspace{-1pt} \alpha \mathcal{L}_{\text{ILM}}(\theta_\text{dec}) \hspace{-3pt} \label{eqn:ilmt_aed}
\end{align}
By minimizing the AED ILMT loss, we maximize the internal LM probability of the E2E training transcripts by updating only the AED decoder while maximizing the conditional probability of the training transcripts given input speech by updating the entire AED model.

The procedure of ILMT with the ILME-based inference for the LM integration with an E2E model is the following
\begin{enumerate}
    \item \label{step:ilmt} Train an E2E model with source-domain audio-transcript pairs to minimize the ILMT loss in Eq. \eqref{eqn:ilmt_rnnt} for RNN-T or in Eq. \eqref{eqn:ilmt_aed} for AED.
    \item \label{step:external_lm} Train an external LM with target-domain text-only data.
    \item Integrate the ILMT E2E model in Step \ref{step:ilmt} with the external LM in Step \ref{step:external_lm} by performing ILME-based inference in Section \ref{sec:ilme}. 
\end{enumerate}

With ILMT, a standalone internal LM with a significantly lower perplexity is learned only by the E2E components used to compute the internal LM scores during the ILME-based inference. With increased modularity, the E2E model is more adaptable to the target domain with its increased flexibility to eradicate the effect of the source-domain internal LM through the ILME-based inference.

\vspace{-0pt}
\section{Experiment}
In this work, we perform ILMT of RNN-T and AED models and integrate them with external long short-term memory (LSTM) \cite{sak2014long, meng2017deep, erdogan2016multi} LMs using different methods. We conduct both cross-domain and intra-domain evaluations to investigate the effectiveness of ILMT. Same as \cite{meng2021ilme}, we perform beam search inference with a beam size of 25 for all evaluations, and use the 3999 
word-piece units generated by byte-pair encoding \cite{sennrich2015neural} as $\mathcal{V}$ for both E2E models and LSTM-LMs. 

\subsection{Internal LM Training}
We perform ILMT of the E2E models with the same 30K hours of anonymized and transcribed data as in \cite{meng2021ilme} collected from Microsoft services, including voice search, short message dictation, command and control, and conversations recorded in various conditions. 

The RNN-T model is initialized with the parameters of the RNN-T in \cite{meng2021ilme} which was well-trained until full convergence with the 30K-hour data. The encoder and prediction networks are both uni-directional LSTMs with 6 and 2 hidden layers, respectively, and 1024 hidden units in each layer.
The joint network has 4000-dimensional output units. 
The RNN-T has 76M parameters. During ILMT, the weight of the internal LM loss is set to 0.4. The internal LM perplexities of ILMT RNN-T and the standard RNN-T in \cite{meng2021ilme} are 52.0 and 99.4, respectively, on the validation set of 30K-hour data.

The AED model \cite{chorowski2015attention, meng2019character, gaur2019acoustic} is randomly initialized and shares the same architecture as the one in \cite{meng2021ilme}. The encoder is a bi-directional LSTM
 with 6 hidden layers and 780 hidden units in each layer. 
The decoder is a uni-directional LSTM with 2 hidden layers, each with 1280 hidden units. The decoder has 4000-dimensional output units.
The AED model has 97M parameters. During ILMT, the weight of the internal LM loss is set to 1.0. The internal LM perplexities of ILMT AED and the standard AED in \cite{meng2021ilme} are 46.1 and 796.7, respectively, on the validation set of the 30K-hour training data.

\subsection{Cross-Domain Evaluation}
\label{sec:cross_eval}
We evaluate a 30K-hour E2E model on the LibriSpeech clean test set by integrating an LSTM-LM trained with LibriSpeech text. Collected from read English based on audio book, the LibriSpeech corpus \cite{panayotov2015librispeech} is outside the domains covered by the 30K-hour training speech. The test-clean and dev-clean sets consist of 2620 and 2703 utterances, respectively. We tune the LM weights on dev-clean. We use the same external LSTM-LM as in \cite{meng2021ilme} trained with the transcript of 960K-hour training speech and the additional 813M-word text from LibriSpeech corpus. With 58M parameters, the LSTM-LM has 2 hidden layers with 2048 hidden units for each layer. For Density Ratio, we use the same source-domain LSTM-LM with 2 hidden layers, 2048 hidden units, and 57M parameters as in \cite{meng2021ilme} trained using the transcript of 30K-hour speech.


We list the results of RNN-T in Table \ref{table:rnnt} with an excerpt of standard E2E training results from \cite{meng2021ilme}.
With ILMT, all three LM integration methods show 27.9\%-40.9\% relative WER reductions from the baseline with standard RNN-T training and inference, significantly larger than the corresponding reductions without ILMT in the range of 16.1\%-29.1\%. 
ILMT with ILME inference performs the best achieving 29.6\% and 16.6\% relative WER reduction from the standard RNN-T with Shallow Fusion and ILME inference, respectively.
As in Table \ref{table:aed}, the AED results are similar to RNN-T. ILMT with ILME inference performs the best, achieving 57.6\%, 31.5\% and 25.1\% relative WER reductions from the standard AED training with AED inference, Shallow Fusion and ILME inference, respectively.

\begin{table*}[h]
\centering
\setlength{\tabcolsep}{5.0pt}
\begin{tabular}[c]{c|c|c||c|c|c||c|c|c||c|c|c}
	\hline
	\hline
	\multirow{3}{*}{\begin{tabular}{@{}c@{}} Train \\ Loss \end{tabular}} & 
	\multirow{3}{*}{\begin{tabular}{@{}c@{}} Evaluation \\ Method \end{tabular}} & \multirow{3}{*}{\begin{tabular}{@{}c@{}} Model \\ Params \end{tabular}} & 
	\multicolumn{3}{c||}{LibriSpeech} & \multicolumn{3}{c||}{Dictation} & \multicolumn{3}{c}{Conversation} \\
	\hhline{~~~---------}
	& & & 
	\multirow{2}{*}{\begin{tabular}{@{}c@{}} Dev \\ WER \end{tabular}} & 
	\multirow{2}{*}{\begin{tabular}{@{}c@{}} Test \\ WER \end{tabular}} & 
	\multirow{2}{*}{\begin{tabular}{@{}c@{}} Test \\ WERR \end{tabular}} &
	\multirow{2}{*}{\begin{tabular}{@{}c@{}} Dev \\ WER \end{tabular}} & 
	\multirow{2}{*}{\begin{tabular}{@{}c@{}} Test \\ WER \end{tabular}} & 
	\multirow{2}{*}{\begin{tabular}{@{}c@{}} Test \\ WERR \end{tabular}} &
	\multirow{2}{*}{\begin{tabular}{@{}c@{}} Dev \\ WER \end{tabular}} & 
	\multirow{2}{*}{\begin{tabular}{@{}c@{}} Test \\ WER \end{tabular}} & 
	\multirow{2}{*}{\begin{tabular}{@{}c@{}} Test \\ WERR \end{tabular}} \\
	& & & & & & & & & & & \\
	\hline
	\multirow{4}{*}{\begin{tabular}{@{}c@{}} $\mathcal{L}_\text{RNN-T}$
		\end{tabular}} & 
	No LM & 76M & 9.27 & 8.97 & - & 23.40 & 16.16 & - & 14.92 & 14.26 & - \\
	\hhline{~-----------}
	& Shallow Fusion & 134M & 7.44 & 7.53 & 16.1 & 22.19 & 15.77 & 2.4 & 14.88 & 14.08 & 1.3 \\
	\hhline{~-----------}
	& Density Ratio & 191M & 6.80 & 6.74 & 24.9 & 21.54 & 15.64 & 3.2 & 14.76 & 14.20 & 0.4 \\
	\hhline{~-----------}
	& ILME & 134M & 6.41 & 6.36 & 29.1 & 21.04 & 14.70 & 9.0 & 14.61 & 14.03	& 1.6  \\
	\hhline{------------}
	\multirow{4}{*}{\begin{tabular}{@{}c@{}} $\mathcal{L}_\text{ILMT}$
		\end{tabular}} & 
	No LM & 76M & 8.58 & 8.37 & 6.7 & 22.61 & 15.73 & 2.7 & 14.00 & 13.59 & 4.7  \\
	\hhline{~-----------}
	& Shallow Fusion & 134M & 6.60 & 6.47 & 27.9 & 21.31 & 15.04 & 6.9 & 13.83 & 13.29 & 6.8 \\
	\hhline{~-----------}
	& Density Ratio & 191M & 5.86 & 5.61 & 37.5 & 20.61 & 14.76 & 8.7 & 13.71 & 13.29 & 6.8 \\
	\hhline{~-----------}
	& ILME & 134M & \textbf{5.57} & \textbf{5.30} & \textbf{40.9} & \textbf{19.94} & \textbf{13.97} & \textbf{13.6} & \textbf{13.32} & \textbf{12.96} & \textbf{9.1} \\
	\hline
	\hline
\end{tabular}
\vspace{-5pt}
  \caption{WERs (\%) of 30k-hour \textbf{RNN-T models} trained with RNN-T or ILMT loss, and evaluated with different LM integration methods on \textbf{out-of-domain} LibriSpeech, \textbf{in-domain} dictation, and \textbf{in-domain} conversation dev/test sets. WERR is relative WER reduction.}
\vspace{-5pt}
\label{table:rnnt}
\end{table*}

\begin{table*}[h]
\centering
\setlength{\tabcolsep}{5.2pt}
\begin{tabular}[c]{c|c|c||c|c|c||c|c|c||c|c|c}
	\hline
	\hline
	\multirow{3}{*}{\begin{tabular}{@{}c@{}} Train \\ Loss \end{tabular}} & 
	\multirow{3}{*}{\begin{tabular}{@{}c@{}} Evaluation \\ Method \end{tabular}} & \multirow{3}{*}{\begin{tabular}{@{}c@{}} Model \\ Params \end{tabular}} & 
	\multicolumn{3}{c||}{LibriSpeech} & \multicolumn{3}{c||}{Dictation} & \multicolumn{3}{c}{Conversation} \\
	\hhline{~~~---------}
	& & & 
	\multirow{2}{*}{\begin{tabular}{@{}c@{}} Dev \\ WER \end{tabular}} & 
	\multirow{2}{*}{\begin{tabular}{@{}c@{}} Test \\ WER \end{tabular}} & 
	\multirow{2}{*}{\begin{tabular}{@{}c@{}} Test \\ WERR \end{tabular}} &
	\multirow{2}{*}{\begin{tabular}{@{}c@{}} Dev \\ WER \end{tabular}} & 
	\multirow{2}{*}{\begin{tabular}{@{}c@{}} Test \\ WER \end{tabular}} & 
	\multirow{2}{*}{\begin{tabular}{@{}c@{}} Test \\ WERR \end{tabular}} &
	\multirow{2}{*}{\begin{tabular}{@{}c@{}} Dev \\ WER \end{tabular}} & 
	\multirow{2}{*}{\begin{tabular}{@{}c@{}} Test \\ WER \end{tabular}} & 
	\multirow{2}{*}{\begin{tabular}{@{}c@{}} Test \\ WERR \end{tabular}} \\
	& & & & & & & & & & & \\
	\hline
	\multirow{4}{*}{\begin{tabular}{@{}c@{}} $\mathcal{L}_\text{AED}$
		\end{tabular}} & 
	No LM & 97M  & 8.56 & 8.61 & - & 20.17 & 14.08 & - & 14.05 & 13.43 & - \\
	\hhline{~-----------}
	& Shallow Fusion & 155M & 5.00 & 5.33 & 38.1 & 18.55 & 12.96 & 8.0 & 13.45 & 12.95 & 3.6  \\
	\hhline{~-----------}
	& Density Ratio & 212M & 4.74 & 5.09 & 40.9 & 18.76 & 12.89 & 8.5 & 13.55 & 12.95 & 3.6 \\
	\hhline{~-----------}
	& ILME & 155M & 4.42 & 4.87 & 43.4 & 18.26 & 12.36 & 12.2 & 13.33 & 12.67 & 5.7 \\
	\hhline{------------}
	\multirow{4}{*}{\begin{tabular}{@{}c@{}} $\mathcal{L}_\text{ILMT}$
		\end{tabular}} & 
	No LM & 97M & 7.31 & 7.47 & 13.2 & 21.06 & 13.72 & 2.6 & 12.60 & 12.19 & 9.2  \\
	\hhline{~-----------}
	& Shallow Fusion & 155M & 6.54 & 6.61 & 23.2 & 19.09 & 12.32 & 12.5 & 12.42 & 11.90 & 11.4 \\
	\hhline{~-----------}
	& Density Ratio & 212M & 4.28 & 4.85 & 43.7 & 18.30 & 12.23 & 13.1 & 12.23 & 11.85 & 11.8 \\
	\hhline{~-----------}
	& ILME & 155M & \textbf{3.30} & \textbf{3.65} & \textbf{57.6} & \textbf{17.00} & \textbf{11.60} & \textbf{17.6} & \textbf{12.11} & \textbf{11.58} & \textbf{13.8} \\
	\hline
	\hline
\end{tabular}
\vspace{-5pt}
  \caption{WERs (\%) of 30k-hour \textbf{AED models} trained with AED or ILMT loss, and evaluated with different LM integration methods on \textbf{out-of-domain} LibriSpeech, \textbf{in-domain} dictation, and \textbf{in-domain} conversation dev/test sets. WERR is relative WER reduction.}
\vspace{-12pt}
\label{table:aed}
\end{table*}

\subsection{Intra-Domain Evaluation}
\label{sec:intra_eval}
\subsubsection{Dictation Test Set}
\label{sec:in_house_dictation}
We evaluate a 30K-hour E2E model on an dictation test set by integrating a strong external LSTM-LM trained with a large amount of multi-domain text. As in \cite{meng2021ilme}, we use the same 2K in-house dictation utterances collected from the keyboard input as the test set, and the same 442 email dictation utterances as the validation set. The test set has a similar style as the dictation data in 30K-hour corpus and is thus considered as in-domain evaluation.
We use the same multi-domain LSTM-LM as in \cite{meng2021ilme} trained with 2 billion (B) words of text comprising short message dictation and conversational data such as talks, interviews, and meeting transcripts. 

Table \ref{table:rnnt} lists the RNN-T results with an excerpt of standard E2E training results from \cite{meng2021ilme}. With ILMT, all three LM integration methods show 6.9\%-13.6\% relative WER reductions from the baseline with standard RNN-T training and inference, significantly larger than the corresponding reductions without ILMT in the range of 2.4\%-9.0\%. 
ILMT with ILME inference performs the best achieving 11.4\% and 5.0\% relative WER reduction from the standard RNN-T training with Shallow Fusion and ILME inference, respectively. 
As in Table \ref{table:aed}, the AED results are similar to RNN-T. ILMT with ILME inference performs the best, achieving 17.6\%, 10.5\% and 6.1\% relative WER reductions from the standard AED training with AED inference, Shallow Fusion and ILME inference, respectively.

\vspace{-5pt}
\subsubsection{Conversation Test Set}
We evaluate a 30K-hour E2E model on an conversation test set by integrating a strong multi-domain external LSTM-LM.
From the Microsoft telecommunication applications, we collect 2560 in-house conversational utterances as the test set, and another 1280 conversational utterances as the validation set. The test set has a similar style as the conversational data in 30K-hour corpus and is thus considered as in-domain evaluation.
For the external LM, we use the same 2B-word LSTM-LM as in Section \ref{sec:in_house_dictation}. 

As shown in Table \ref{table:rnnt}, with ILMT RNN-T, all three LM integration methods show 6.8\%-9.1\% relative WER reductions from the baseline with standard RNN-T training and inference, significantly larger than the corresponding reductions without ILMT in the range of 0.4\%-1.6\%. 
ILMT with ILME inference performs the best achieving 8.0\% and 7.6\% relative WER reduction from the standard RNN-T training with Shallow Fusion and ILME inference, respectively.
As shown in Table \ref{table:aed}, the AED results are similar to RNN-T. ILMT with ILME inference performs the best, achieving 13.8\%, 10.6\% and 8.6\% relative WER reductions from the standard AED training with AED inference, Shallow Fusion and ILME inference, respectively.

\vspace{-4pt}
\subsection{Result Analysis}
From the results, 
we have the following observations for both RNN-T and AED models, and for both cross- and intra-domain evaluations. All LM integration methods consistently achieve remarkably lower WERs with ILMT than with standard E2E training. 
Among all methods, ILMT with ILME inference consistently performs the best, with 29.6\%-31.5\% and 8.0\%-11.4\% relative WER reductions from standard E2E training with Shallow Fusion for cross-domain and intra-domain evaluations, respectively.
ILME inference consistently outperforms Density Ratio in terms of lower WER with ILMT or standard E2E training despite having 26.8\%-29.8\% fewer model parameters.
With ILME inference, ILMT achieves 16.6\%-25.1\% and 5.0\%-8.6\% relative WER reductions from the standard E2E training for cross-domain and intra-domain evaluations, respectively.
All of these manifest the advantage of ILMT over standard E2E training for ILME inference and other LM integration methods.

Note that, with or without ILMT, ILME inference is effective even for intra-domain evaluation because it replaces the weak E2E internal LM with a powerful external LM trained with orders of magnitude more multi-domain text than the E2E training transcript. 
All three LM fusion methods perform better for AED than RNN-T due to larger relative WER reductions from a stronger baseline.
The internal LM perplexity of an E2E model is remarkably reduced by ILMT. 


\vspace{-0pt}
\section{Conclusion}
We propose an internal LM training of the E2E model which minimizes an internal LM loss in addition to the standard E2E loss to improve the effectiveness of ILME external LM integration. With ILMT, ILME inference achieves 29.6\%-31.5\% and 8.0\%-11.4\% relative WER reductions from the standard E2E training with Shallow Fusion for cross-domain and intra-domain evaluations, respectively. With ILME inference, ILMT outperforms the standard E2E training by 16.6\%-25.1\% and 5.0\%-8.6\% relatively in terms of lower WER for cross-domain and intra-domain evaluations, respectively.


\vfill\pagebreak

\bibliographystyle{IEEEbib}
\bibliography{strings,refs}

\end{document}